\documentclass{ws-ijmpcs}
\usepackage{amssymb,amsfonts,amsmath}
\usepackage{graphicx}
\usepackage{mathptmx}
\newcommand{\HeII}{He-II }
\newcommand{\HeI}{He-I }
\newcommand{\He}{${}^4$He~}

\begin{document}

\title{Quantum fluids in nanoporous media - effects of the confinement and
fractal geometry}

\author{D.A. Tayurskii}

\address{Department of Physics , Kazan Federal University, 18 Kremlevskaya st.\\
Kazan, 420008, Russia\\
dtayursk@gmail.com}

\author{Y.V. Lysogorskiy}

\address{Department of Physics , Kazan Federal University, 18 Kremlevskaya st.\\
Kazan, 420008, Russia\\
yurii.lysogorskii@gmail.com}

\maketitle
\begin{abstract}
The zero-point energy of atoms of two stable isotopes of helium -
${}^4$He and ${}^3$He - is high enough to prevent their
solidification even at extremely low temperatures without
application of external pressure. So they belong to the class of
quantum fluids with strong correlations between atoms but their
behavior is quite different at low temperatures. The first one
represents a Bose-system and shows superfluid transition at 2.17 K
while atoms of 3He are governed by Fermi-statistics and superfluid
transition can be observed only at much lower temperatures (about
1 mK) when the pairing of two atoms occurs. But even at helium
temperatures (1.5-4.2 K) the effects of quantum statistics for
${}^3$He atoms becomes pronounceable especially in nanoscale
confinement (nanoporous media, thin adsorbed layers on solid
substrates) and in the presence of nanoscale disorder induced for
example by silica aerogel strands. In recent years the problem of
correct description of quantum fluids in the confined geometry at
nanoscale length has emerged. It has been recognized that the
quantum fluids at these circumstances can be considered as a new
state of quantum matter due to close values between characteristic
lengths for these quantum liquids and the size of geometrical
confinement and significant contribution from the surface atoms.
So one has to apply new physics to describe such systems with
taking into account their complex nature. For example, last two
years the attempts to develop the fractionalized two-fluid
hydrodynamics for nanoporous media with fractal dimensions have
been made. The actuality of such new hydrodynamics becomes very
clear for the last development in chemical synthesis of different
kind of aerogels with nanopore structure as well as numerous
studies of nanoporous substances. One of the interesting obtained
results that density waves (the first sound) and temperature waves
(the second sound) become strong coupled even in the absence of
viscosity, so it is purely geometric effect of fractal space of
nanopores. In the present report we will review the procedure,
results and discuss the issues for this approach.\\
 This work is supported in part by the Russian Fund for the Fundamental Research
(09-02-01253) and by the Ministry of Education and Science of the
Russian Federation (FTP "Scientific and scientific-pedagogical
personnel of the innovative Russia" contract N 02.740.11.0797)
\end{abstract}

\section{Introduction}
The zero-point energy of atoms of two stable isotopes of helium -
${}^4$He and ${}^3$He - is high enough to prevent their
solidification even at extremely low temperatures without
application of external pressure. So they belong to the class of
quantum fluids with strong correlations between atoms but their
behavior is quite different at low temperatures. The first one
represents a Bose-system and shows superfluid transition at 2.17 K
while atoms of 3He are governed by Fermi-statistics and superfluid
transition can be observed only at much lower temperatures (about
1 mK) when the pairing of two atoms occurs. But even at helium
temperatures (1.5-4.2 K) the effects of quantum statistics for
${}^3$He atoms becomes pronounceable especially in nanoscale
confinement (nanoporous media, thin adsorbed layers on solid
substrates) and in the presence of nanoscale disorder induced for
example by silica aerogel strands. In recent years the problem of
correct description of quantum fluids in the confined geometry at
nanoscale length has emerged\cite{Wong,Matsumoto,Vasquez,Azuah}.
It has been recognized that the quantum fluids at these
circumstances can be considered as a new state of quantum matter
due to close values between characteristic lengths for these
quantum liquids and the size of geometrical confinement and
significant contribution from the surface atoms. So one has to
apply new physics to describe such systems with taking into
account their complex nature. For example, last two years the
attempts to develop the fractionalized two-fluid hydrodynamics for
nanoporous media with fractal dimensions have been
made\cite{TayuLys1,TayuLys2}. The actuality of such new
hydrodynamics becomes very clear for the last development in
chemical synthesis of different kind of aerogels with nanopore
structure as well as numerous studies of nanoporous substances.
One of the interesting obtained results that density waves (the
first sound) and temperature waves (the second sound) become
strong coupled even in the absence of viscosity, so it is purely
geometric effect of fractal space of nanopores.

In the present report we will review the procedure, results and
discuss the issues for this approach. The rest of this paper is
organized as follow: In Section~\ref{SEC_existed_model} we make a
little review of existed model of superfluid  bulk helium-4. In
the next Section~\ref{SEC_aerogel} propeties and peculiarities of
aerogel are discussed. In Section~\ref{SEC_fractTFM} possible
fractionalization of hydrodynamical model is shown. In
Section~\ref{SEC_fractNLS} we speculate about another way of
considering complex fractal structure of aerogel for description
of superfluid helium behavior.  In Section~\ref{SEC_conclusion} we
make a summarization and some novel results are given.


\section{Existed model}
\label{SEC_existed_model} There are several approaches to describe
the behavior of bulk superfluid helium-4. For example, two-fluid
model\cite{Khalatnikov,Tilley}, the microscopic description based
on the Gross-Pitaevskii equation\cite{Coste,HillsRoberts} and
others.

\subsection{Two-fluid model}
The most famous model is two-fluid model (TFM) proposed by Landau
in the past century. In this model superfluid helium with density
$\rho$ is considered as a two component system: an uncondensed,
normal component with density $\rho_n$ with velocity $v_n$ and a
condensed, superfluid component characterized by density $\rho_s
=\rho-\rho_n$ with velocity $v_s$. Without dissipative terms one
finds the following system of hydrodynamical equations (so-called
Landau-Khalatnikov equations\cite{Khalatnikov,Tilley} )

\begin{eqnarray}
\label{TFrho} \frac{d \rho}{d t}&+& div( \rho_n\textbf{v}_n+\rho_s\textbf{v}_s ) = 0,\\
\label{Tfj} \frac{\partial}{\partial t} (
\rho_n\textbf{v}_n+\rho_s\textbf{v}_s )_i&+&
\frac{\partial}{\partial x_k}\Pi_{ik}=0,\\
\label{TFS}\frac{d S}{d t}&+&div S \textbf{v}_n=0,\\
\label{TFvs} m\frac{\partial\textbf{v}_s}{\partial
t}&+&m(\textbf{v}_s\nabla)\textbf{v}_s +\nabla\mu=0,
\end{eqnarray}
The entropy is denoted by $S$, the mass of an atom by $m$, the
pressure by $p$ and the chemical potential by $\mu$.
In~(\ref{Tfj}) the stress tensor $\Pi_{ik}$ is given by
\begin{equation}
\label{Piik1} \Pi_{ik}=\rho_n v_{ni} v_{nk}+ \rho_s v_{si}
v_{sk}+p \delta_{ik},
\end{equation}
and the sum over the index $k$ is assumed.

One can obtain from this equations that there are two type of
collective motion: motion of the fluid where both components move
in phase is called first (ordinary) sound, while second sound is
associated with out of phase motion of the two components. The
above TFM equations describe the flow properties of superfluid
$^4$He in the bulk of the liquid. They are not valid in the
immediate proximity of walls and free surfaces, where the effects
of so-called "healing" are encountered. Moreover,
Landau-Khalatnikov equations cannot be used in the vicinity of
$\lambda$ point, where large variations of superfluid density in
space as well as in time may occur.

For taking into account such "healing" effects, which has quantum
nature, one can consider some models derived from nonlinear
Schr\"{o}dinger equations.

\subsection{Nonlinear Schr\"{o}dinger equation}
The Schr\"{o}dinger equation itself is applicable only at $T=0$K
and small interaction between helium atoms, when all of them are
in condensed state. In that case we can introduce macroscopic wave
function of condensate. Than velocity field is determined as
probability flow of this wave function and it is proportional to
phase gradient. He-II is dense fluid of strongly interacted bosons
and applicability of NLS is questionable, but it helps us to
elucidate basic peculiarities of superfluid dynamics.

If one writes short range pair interaction between particles in
form $U(x-x^\prime)=U_0\delta(x-x^\prime)$,
$N_0=\int\left|\Psi\right|^2d^3x$ - number of particles in
condensate and $V_0$ - total condensate volume, than one can write
\textit{nonlinear Schr\"{o}dinger equation}\cite{Coste} in form of
\begin{equation}
\label{NLS}
 i\frac{\partial \Psi}{\partial
t}=-\frac{1}{2}\nabla^2\Psi+\left|\Psi\right|^2-\Psi.
\end{equation}

Here average density $\rho_0=N_0/V_0$ is set to unity, unity of
length is $\hbar/\sqrt{\rho_mU_0}$ and unity of time
$\hbar/(\rho_0 U_0)$.

After substitution $\Psi(x,t)\equiv\sqrt{\rho(x,t)}e^{i
\theta(x,t)}$ and disparting real and imaginary parts in
(\ref{NLS}) one can obtain
 \begin{eqnarray}
 \label{NLSrho} \frac{ \partial \rho}{\partial t} &+& \nabla\cdot\rho v=0\\
\label{NLSv} \frac{\partial v}{\partial t}&+&\nabla
\frac{v^2}{2}+\frac{1}{\rho}\nabla\frac{\rho^2}{2}=-\nabla\left(\frac{(\nabla
\rho)^2}{8\rho^2}-\frac{\nabla^2\rho}{4\rho}\right),
 \end{eqnarray}
where $v\equiv\nabla \theta$. Except for high-order derivatives in
right hand side of equation~(\ref{NLSv}), which can be omitted in
hydrodynamical limit, system of
equations~(\ref{NLSrho})-(\ref{NLSv}) is equivalent to Euler
equation for nonrotating ideal fluid with pressure defined as
$p(\rho) \equiv \rho^2/2$. Pressure depends only on density $\rho$
because fluid is considered at absolute zero.

For considering dynamics of He-II at nonzero temperature, Hills
and Roberts have improved TFM\cite{HillsRoberts} by introducing
additional terms, that is proportional to the gradient of
superfluid component. In that case equation~(\ref{TFvs}) can be
rewritten as
\begin{equation}
\label{HRvs} \frac{\partial \mathbf{v}_s}{\partial
t}+\nabla\left(\frac{\mathbf{v}_s^2}{2}+\mu\right)=\nabla\left(\eta(\rho_s)\nabla^2\rho_s+\frac{1}{2}\frac{d
\eta}{d \rho_s}(\nabla \rho_s)^2\right),
\end{equation}
where $\eta(\rho_s)$ - some undefined function; stress tensor has
to be modified and rewritten as
\begin{equation}
\label{HLPij} \Pi_{ij} \equiv \rho_n v_{ni} v_{nj} + \rho_s v_{si}
v_{sj}+\eta \nabla_i \rho_s \nabla_j \rho_s + \left(p-\eta \rho_s
\nabla^2 \rho_s-\frac{1}{2} \frac{\rho_s \eta}{\rho_s}(\nabla
\rho_s)^2\right)\delta_{ij}.
\end{equation}
New terms that include spatial derivatives of $\rho_s$ are
responsible for macroscopic quantum effects, like healing
length\cite{HillsRoberts}.

\subsection{Correlated density matrix theory}
Another way for describing quantum strongly correlated system
(superfluid helium is just one example of such systems) is
\textit{correlated density matrix}\cite{CDM} theory that provides
the method of choice to analyze the microscopic structure of
strongly correlated quantum fluids in thermal equilibrium. The
main point of this theory is concept of renormalized bosons and
fermions. Their mass can depend fully on temperature, density and
wave vector. So one may discuss the properties of strongly
correlated quantum fluids at finite temperatures with reference to
the background gas of renormalized free particles.

\subsection{Extended irreversible thermodynamics}
Recently the \textit{extended irreversible thermodynamics} has
been applied to describe \HeI and \HeII\cite{Mongiovi0,Mongiovi}.
It is shown that the behavior of helium II can be described by
means of an extended thermodynamic theory where four fields,
namely density, temperature, velocity, and heat flux are involved
as independent fields. This model is able to explain the
propagation of the two sounds that are typical of helium II, and
the attenuation calculated for such sounds is in agreement with
the experiment results. After some approximations, this model is
reduced to TFM, so it can be considered as some generalization of
TFM.

\section{Aerogel}
\label{SEC_aerogel}
 Silica aerogels are synthesized via a sol-gel
process and hypercritical drying which enable production of
tenuous solids with porosity $\phi$ as large as 99.8 \% and unique
acoustic properties. Silica aerogel are known to be good examples
of fractal materials. A revealed by small-angle x-ray-scattering
(SAXS) experiments or small-angle neutron-scattering (SANS)
experiments, they are made of a disorder, but homogeneous, array
of connected fractal clusters resulting from the aggregation of
primary particles. The analysis of the wave-vector dependence of
the scattering intensity $I(q)$ has permitted the determination of
two characteristic length scales which are the average size
$a\approx 10$~\AA~of the particles and the average size of the
clusters $\xi \approx 100$~\AA. At length scales from $a$ to $\xi$
silica aerogels show a fractal behavior\cite{Porto}.

The computational confirmation for cluster structure of aerogel
has been obtained by modeling as well as by the geometrical
analysis of the diffusion limited cluster-cluster
aggregation\cite{Courtens,Hasmy}.

 Also it has been demonstrated\cite{Vasquez} that it is
long-correlated structure of aerogel that makes an essential
influence on liquid \He behavior near $\lambda$-point.



\subsection{Behavior of \He inside aerogel}
There has been considerable interest in the behavior of superfluid
\He in the presence of a random disorder induced by highly open
porous media, like aerogel.  Understanding the results of acoustic
experiments is important when dealing with porous media. Use of
liquid \He offers unique advantages due to the existence of the
superfluid phase with more than one sound mode. In a porous media
where the normal component is clamped by its viscosity and only
the superfluid component can move, fourth sound (relative motion
of the superfluid and normal fluids) propagates and can be used to
determine the superfluid fraction.

The high-porosity aerogels are so soft that the aerogel matrix and
the clamped normal fluid moves as the results of pressure and
temperature gradients, unlike other porous media. This results in
sound modes intermediate between first and fourth sound and a
second-sound-like mode. In that case, proposed by Biot theory of
acoustic propagation in porous, fluid filled, macroscopically
homogeneous and isotropic media, is not applicable.

It is very interesting to study the possible influence of
geometrical confinement with fractal dimensionality on the flow
properties of superfluid $^4$He in the framework of TFM. In order
to shed a light on geometrical factor itself one can neglect here
by quantum healing and any dissipative processes.

\subsection{Nonextensivity}

 So hereafter we suppose that effectively aerogel can be
considered as a cluster with a fractal mass dimension\cite{Feder}
the nanopores of which are filled in by liquid helium.

Note that thermodynamic limit conditions are violated for helium
atoms inside nanopores because of a huge inner pore surface of
aerogel (up to 3000 m$^2$/g for an aerogel with density 2
mg/cm$^3$). Namely the ratio of total number of helium atoms $N$
to the total cluster volume $V$ is not constant at $N,V
\rightarrow \infty$ and nonextensivity of physical properties
takes place for helium atoms in two nanopores. In this case the
methods of non-extensive thermodynamics\cite{NE} should be applied
to construct the two-fluid hydrodynamic model and non-extensive
entropy like Tsallis entropy should be introduced. We have to note
here that beyond thermodynamic limit even the usual
Boltzmann-Gibbs entropy becomes non-additive, but the additivity
is restored when thermodynamic limit conditions are taken
\cite{Parvan}.

\section{Fractionalized and nonextensive hydrodynamical model}
\label{SEC_fractTFM} In general case a macroscopic quantity
$Q(A,B)$ associated with the total system may be expressed in
terms of the same quantity associated with the subsystems, $Q(A)$
and $Q(B)$~\cite{Badescu}
\begin{equation}
\label{generalcomposfunc}
 Q(A,B)= f_{\lambda_Q}[Q(A),Q(B)],
\end{equation}
where $f_{\lambda_Q}$ is a symmetric bivariate function depending
on a constant $\lambda_Q$. Of course, for given quantity $Q$ there
exist many functions which satisfy the composability
property~(\ref{generalcomposfunc}). However, additional
assumptions drastically reduce their number. For example, the
thermodynamic equilibrium may be used as a constraint on the form
of $f_{\lambda_Q}$ in Eq.~(\ref{generalcomposfunc}). For energy
$E_\lambda$\cite{QAWang} and entropy $S_q$\cite{SAbe} of helium
inside nanopores it leads to
\begin{eqnarray}
\label{NEE} E_\lambda(A+B)&=&E_\lambda(A)+E_\lambda(B)+\lambda E_\lambda(A) E_\lambda(B),\\
\label{NES} S_q(A+B)&=&S_q(A)+S_q(B)+q S_q(A) S_q(B),
\end{eqnarray}
where $\lambda$ and $q$ are parameters of nonextensivity and are
determined by the properties of system. Suppose that a local
equilibrium between liquid $^4$He in different nanopores takes
place. So we can re-define such thermodynamics quantities as
temperature and pressure in form of\cite{Abe}.
\begin{eqnarray}
\label{Tphys} T_{phys}&=&\frac{1+q S_q}{1+\lambda E_\lambda}
\left(\frac{\partial E_\lambda}{\partial S_q}\right)_V,\\
\label{pphys} p_{phys}&=&\frac{T_{phys}}{1+q
S_q}\left(\frac{\partial S_q}{\partial V}\right)_E.
\end{eqnarray}
Further, it is possible to introduce the spatial pressure and
density distributions in fractal cluster as it has been made in
our previous work\cite{TayuLys1}
\begin{eqnarray}
\label{pphys1} p_{phys}(r)&=&p_f(r) \chi_p(r),\\
\label{rhophys} \rho_{phys}(r)&=&\rho_f(r) \chi_\rho(r),
\end{eqnarray}
where  $r$ is distance from center of fractal cluster, $p_f(r)$ is
distribution of pressure in pore, $\chi_p(r)$ is a fractal
factor-function. In the case of Euclidean space with $D=3$  this
factor-function should be equal to unit: $\chi_p(r)\equiv 1$. Such
kind of fractionalization procedure can be applied to any
thermodynamic quantity $A$ i.e.
\begin{eqnarray}
&A=A_f(X,Y,Z,...) \chi_A (r),\\
&\lim_{D\rightarrow 3} \chi_A=1.
\end{eqnarray}

\begin{figure}
\begin{center}
\includegraphics[%
  width=0.7\linewidth, keepaspectratio]{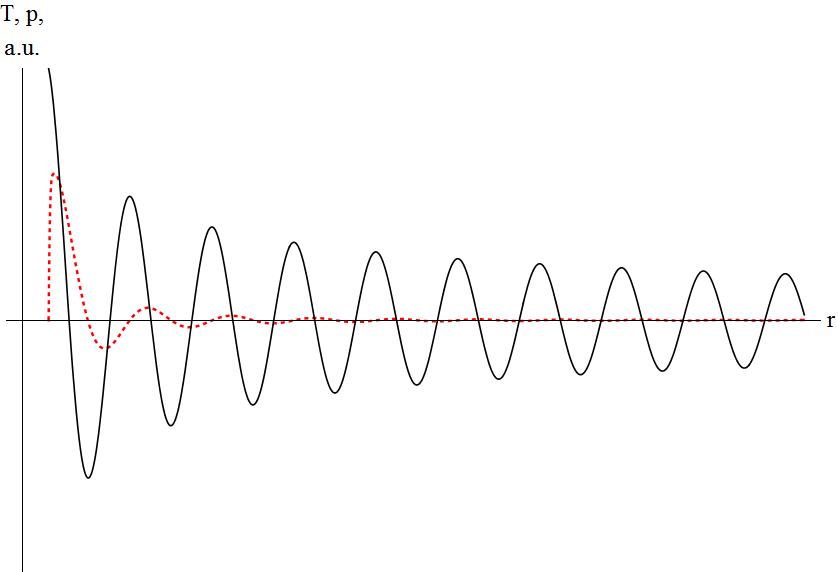}
\end{center}
\caption{Profile for pressure wave (black line) and the
temperature wave induced by it(dashed line).}
 \label{TnP}
\end{figure}

From the defined thermodynamic
quantities~(\ref{Tphys},\ref{pphys}) and eqn.~(\ref{pphys1}) the
fractal factor-functions for energy, entropy and temperature are
derived
\begin{eqnarray}
\chi_E&=&\chi_S=\frac{\chi_p}{1+\lambda E_f (1-\chi_p)},\\
\chi_T&=&\frac{1+q S_f\chi_S}{1+ q S_f}\frac{1+\lambda
E_f}{1+\lambda E_f \chi_E}.
\end{eqnarray}
Because all thermodynamic quantities should be expressed in terms
of physical (observable) variables, one can propose the following
definition of the generalized free energy
\begin{equation}
\label{F} F=E-T_{phys} S_{phys}=E-\left(\frac{\partial E}{\partial
S_{phys}}\right) S_{phys}.
\end{equation}
which represents no more than the Legendre transformation. So the
entropy differential equals to
\begin{equation}
dS_{phys}=\frac{1+\lambda E_\lambda}{1+q S_q}dS_q,
\end{equation}
In the first order with respect to $\lambda E_f$ and $q S_q$ the
entropy can be written as:
\begin{equation}
\label{Sphys} S_{phys} = \frac{1}{q} \ln(1+q S_q)+\lambda \int
\frac{E_\lambda(S_q) dS_q}{1+q S_q}\approx S_q-q S_q
\frac{S_q}{2}+\lambda \int E_\lambda dS_q.
\end{equation}
Finally the fractal factor-function for $S_{phys}$ is
\begin{equation}
\chi_{Sp}=\chi_S+\chi_p(\chi_p-1)\left(\frac{\lambda E_f-q
S_f}{2}+\lambda H\right),
\end{equation}
where $H=1/S_f \int E_f dS_f$. After substitution the
fractionalized thermodynamical quantities into TFM one can derive
the main system of equations for the \textit{fractionalized
twofluid hydrodynamic model}
\begin{eqnarray}
\label{FTFMrho}\frac{\partial \rho_f}{\partial t} \chi_p + div
(\rho_{sf} \chi_p
\textbf{v}_s+\rho_{nf}\chi_p \textbf{v}_n)=0,\\
\label{FTFMS}\frac{\partial \rho_f \sigma_f}{\partial t}
\chi_{Sp}+ div (\rho_f
\sigma_f \chi_{Sp} \textbf{v}_n)=0,\\
\label{FTFMvs}\rho_{sf} \chi_p \frac{\partial
\textbf{v}_s}{\partial t}=-\frac{\rho_{sf}}{\rho_f}\nabla(p_f
\chi_p)+\rho_{sf} \sigma_f
\chi_p \chi_\sigma\nabla(T_f \chi_T),\\
\label{FTFMvn}\rho_{nf} \chi_p \frac{\partial
\textbf{v}_n}{\partial t}=-\frac{\rho_{nf}}{\rho_f}\nabla(p_f
\chi_p)-\rho_{sf} \sigma_f \chi_p \chi_\sigma \nabla(T_f \chi_T),
\end{eqnarray}
where $\chi_\sigma = 1+(\chi_p-1)\left(\frac{3\lambda E_f-q
S_f}{2}+\lambda H\right)$. From
eqns.~(\ref{FTFMrho},\ref{FTFMS},\ref{FTFMvs},\ref{FTFMvn})  two
equations for waves of pressure and temperature follow
\begin{eqnarray}
\label{posc} \frac{1}{u_1^2} \frac{\partial^2 p_f}{\partial
t^2}&=&\nabla^2 p_f+2 \frac{\nabla \chi_p}{\chi_p}\nabla
p_f+\frac{\nabla^2
\chi_p}{\chi_p} p_f,\\
\nonumber \frac{1}{u_2^2}\frac{\partial^2 T_f}{\partial
t^2}&=&\nabla^2 T_f \left(1+(\chi_p-1)M\right)+\nabla T_f
\left(M\nabla \chi_p-(1-M)\frac{\nabla
\chi_p}{\chi_p}\right)+\\
\nonumber &+&T_f (q S_f-\lambda E_f) \left(\nabla^2
\chi-\frac{(\nabla \chi_p)^2}{\chi_p}\right) +\\
\label{Tosc} &+&\frac{1}{u^2_2}\frac{N
\sigma_f}{\rho_f}\left(\frac{\partial T_f}{\partial
\sigma_f}\right)_{\rho_f}\left(\nabla \chi_p \nabla p_f +
\frac{(\nabla \chi_p)^2}{\chi_p}p_f\right),
\end{eqnarray}
where $u_1^2=(\partial p_f/\partial \rho_f)_{S_f}$,
$u_2^2=\rho_{sf} \sigma_f^2/\rho_{nf}(\partial T_f/\partial
\sigma_f)_{p_f}$ are the squared first and second sound velocities
respectively and $M=(\lambda E_f+q S_f)/2 +\lambda H$, $N=(3
\lambda E_f-q S_f)/2 +\lambda H$. On figure~\ref{TnP} the profiles
for pressure wave and for the temperature wave induced by it are
shown. It is seen from eqn.~(\ref{Tosc}) that the coupling between
pressure and temperature waves appears even in the absence of
$^4$He viscosity and aerogel sceleton inertia, which is
undoubtedly the effect of not only fractional dimensionality of
nanopore space, but also of the nonextensive nature of
thermodynamical quantities for He-II inside nanopores.

\section{Fractional nonlinear Schr\"{o}dinger equation}
\label{SEC_fractNLS} Let us assume that confinement (aerogel
strands) forbid realization of some Feynman's pathes for
particular helium atom in contrast to helium atom in the free
space. For taking into account such peculiarities one can propose
to make use of fractional Schr\"{o}dinger equation\cite{Laskin}.

Fractional Schr\"{o}dinger equation (FSE) is derived from path
integration over Levy's pathes and is given by
\begin{equation}
\label{FSE}
 i\hbar\frac{\partial \psi(\textbf{r},t)}{\partial t}=K_\alpha
\left(-\hbar^2 \nabla^2\right)^{\alpha/2}
\psi(\textbf{r},t)+V(\textbf{r},t)\psi(\textbf{r},t).
\end{equation}
Here Riesz fractional derivative is introduced as
\begin{eqnarray}
\left(-\hbar^2 \nabla^2\right)^{\alpha/2}
\psi(\textbf{r},t)&=&\frac{1}{(2\pi\hbar)^3}\int d^3pe^{i p
r/\hbar}\left|\textbf{p}\right|^\alpha \varphi(\textbf{p},t),\\
\varphi(\textbf{p},t)&=&\int d\textbf{r} e^{-i p r/\hbar}
\psi(\textbf{r},t).
\end{eqnarray}

\subsection{Galilean noninvariance of fractional Schr\"{o}dinger equation}
Let us consider tranformation of equation~(\ref{FSE}) when
inertial reference frame is changed. With such transformation
spatial coordinates and time are changed as $t^\prime = t$,
$r^\prime=r-v t$, where $v$ is some constant relative velocity.
Then time and fractional spatial derivative is rewritten
as\cite{Osler}
\begin{eqnarray}
\partial_t f &=& \left(\partial_{t^{\prime}}-v
\partial_{r^\prime}\right)f,\\
\left({\nabla_{r}}^2\right)^{\alpha/2} &=& \left(
{\nabla_{r^\prime}}^2\right)^{\alpha/2}.
\end{eqnarray}
Let us assume that in old reference frame \textit{$K$} FSE has
form
\begin{equation}
\label{FSEoldRF} i\hbar\frac{\partial \Psi(\textbf{r},t)}{\partial
t}=K_\alpha \left(-\hbar^2 \nabla^2\right)^{\alpha/2}
\Psi(\textbf{r},t).
\end{equation}
While changing reference frame to \textit{$K^\prime$}, old wave
function is written as\cite{Greenberger}
\begin{equation}
\label{OldWaveFunction} \Psi(\textbf{r},t) =
\varphi(\textbf{r}^\prime,t^\prime) e^{i
f(\textbf{r}^\prime,t^\prime)},
\end{equation}
where $\varphi$ denote wave function in new reference frame.
Fractional derivative of product is revealed by generalized
Leibniz rule:
\begin{equation}
\label{FDProd} D_x^{\alpha} \left(\varphi e^{i
f}\right)=\sum_{n=0}^{\infty}
\frac{\Gamma(\alpha+1)}{\Gamma(\alpha-n+1)\Gamma(n+1)}
D_x^{\alpha-n}\varphi D_x^n e^{i f}.
\end{equation}
After substitution eqn.(\ref{OldWaveFunction}) in (\ref{FSEoldRF})
with taking into account eqn. (\ref{FDProd}) one can derive
\begin{eqnarray}
\nonumber \left(i \hbar \partial_{t^\prime}\varphi-\hbar\varphi
\partial_{t^\prime} f - i \hbar v \nabla_{r^{\prime}}\varphi+\hbar v \varphi \nabla_{r^{\prime}} f\right)e^{i
f} =\\
\label{FSEnewRF} K_\alpha (-\hbar^2)^{\alpha/2}
\sum_{n=0}^{\infty}
\frac{\Gamma(\alpha+1)}{\Gamma(\alpha-n+1)\Gamma(n+1)}
D_{r^{\prime}}^{\alpha-n}\varphi D_{r^{\prime}}^n e^{i f}.
\end{eqnarray}
For Galilean invariance of eqn.~(\ref{FSEnewRF}) one has to keep
only time derivative and spatial fractional derivative of order
$\alpha$. Let us equate to zero coefficient near the rest
derivatives of wave function $\varphi(\textbf{r},t)$. Then we can
obtain that $v=0, \partial_{t^\prime} f = 0, \nabla_{r^\prime} f =
0$, that is $f(r^\prime,t^\prime)=const$. It leads to
\begin{eqnarray}
\nonumber r^\prime &=& r,\\
\nonumber t^\prime &=& t,\\
\nonumber \Psi^\prime &=& \Psi e^{i \lambda},
\end{eqnarray}
where $\lambda$ is some constant. Thus, fractional Schr\"{o}dinger
equation is \textit{Galilean noninvariant}, that is it changes its
form with changing inertial reference frame. As a consequence one
needs to choose the especial reference frame where, for example,
nanoporous media is in a rest as well as to introduce additional
potential into two-fluid hydrodynamic model.

\section{Conclusion}
\label{SEC_conclusion}

It was shown that fractionalized set of hydrodynamical equations
with taking into account nonextensivity of He-II inside nanopores
leads to coupling between the first and the second sounds that
appears even in the absence of viscous friction, which is
undoubtedly the effect of fractional dimensionality of nanopore
space and nonextensive nature of helium droplets. It was proposed
that for better description of behavior of superfluid in
nanoporous media with complex fractal structure one can use
fractional Schr\"{o}dinger equation. It leads to Galilean
noninvariance and as a consequence one needs to choose the
especial reference frame where, for example, nanoporous media is
in a rest as well as to improve two-fluid hydrodynamic model by
introducing there additional potential.

\section*{Acknowledgements}
 This work is supported in part by the Russian Fund for the Fundamental Research
(09-02-01253) and by the Ministry of Education and Science of the
Russian Federation (FTP "Scientific and scientific-pedagogical
personnel of the innovative Russia" contract N 02.740.11.0797)

\end{document}